\documentclass[pra,showpacs,twocolumn,floatfix]{revtex4}
\usepackage{epsfig,psfrag}
\usepackage{amsmath}
\usepackage{textcomp}
\allowdisplaybreaks

\begin{document}

\title{Cooperative fluorescence effects for dipole-dipole interacting
  systems with experimentally relevant level configurations}
\author{Volker Hannstein} 
\author{Gerhard C.~Hegerfeldt}
\affiliation{Institut f\"ur Theoretische Physik, Universit\"at G\"ottingen,
  Tammannstr. 1, 37077 G\"ottingen, Germany}

\begin{abstract}
The mutual dipole-dipole interaction of atoms in a trap can affect
their fluorescence. Extremely large effects were reported for double
jumps between different intensity periods in experiments with two and
three Ba$^+$ ions  for distances in the range of about ten wave lengths of
the strong transition while no effects were observed for Hg$^+$ at 15
wave lengths. 
In this theoretical paper 
we study this question for  configurations with three and four levels
which model those of Hg$^+$ and Ba$^+$, respectively. For two systems
in the Hg$^+$ configuration we find cooperative effects of up to
30\% for distances around one or two wave lengths, about 5\% around
ten wave lengths, and, for larger distances in agreement with
experiments, practically none. This is similar for two V systems. However, for
two four-level configurations, which model two  Ba$^+$ ions, 
cooperative effects are practically absent, and this latter result
is at odds with the experimental findings for Ba$^+$.
\end{abstract}

\maketitle

\section{Introduction}
The dipole-dipole interaction is ubiquitous in physics and for example
responsible for the ever present van der Waals force. It is also
important for envisaged quantum computers based on atoms or ions in
traps. Considerable interest in the literature has also been aroused
by its cooperative effects on the radiative behavior of atoms
\cite{refs:AdBeDaHe}. In an as yet unexplained experiment
\cite{SaBlNeTo:86,Sa:86} 
with two and three Ba$^+$ ions, which exhibit macroscopic light and
dark periods, a large fraction of double and triple jumps was
reported, i.e. jumps by two or 
three intensity steps within a short resolution time. This fraction was
orders of magnitudes larger than for independent ions. The
quantitative explanation of such a large cooperative effect for
distances of the order of ten wave lengths of the strong transition has
been found difficult
\cite{HeNi:88,LeJa:87,LeJa:88,AgLaSo:88,LaLaJa:89,FuGo:92}.
Experiments with other ions showed no observable cooperative effects
\cite{ThBaDhSeWi:92,BeRaTa:03}, 
in particular none were seen for Hg$^+$ for a distance of about 15 wave lengths
\cite{ItBeWi:88}. More recently, an unexpected high number of simultaneous
quantum jumps in a linear chain of trapped Ca$^+$ ions was reported
\cite{BlReSeWe:99} while no such effects were
found in another experiment \cite{DoLuBaDoStStStSt:00} using the same
ion species and a similar setup. 
 
Systems with macroscopic light and dark periods can provide a
sensitive test for cooperative effects of the dipole-dipole
interaction. These periods can occur for multi-level systems where the
electron is essentially shelved for some time in a meta-stable state
without photon emission
\cite{ref:BeHe}. For two V systems with macroscopic
light and dark periods the effect of the dipole-dipole interaction was
investigated numerically in Ref.~\cite{BeHe:99} and analytically in Ref.~\cite{AdBeDaHe:01} and shown to be up to 30\% in the
double jump rate compared to independent systems. Monitoring the
dipole-dipole interaction of two V systems via quantum jumps of
individual atoms was investigated in Ref.~\cite{SkoZaAgWeWa:01}. The
experimental systems of Refs.~\cite{SaNeBlTo:86,SaBlNeTo:86,ItBeWi:88}
are, however, not in the V configuration so that the results of
Ref.~\cite{BeHe:99} do not directly apply.

\begin{figure}[b]
  \psfrag{1}{$|1\rangle$}
  \psfrag{2}{$|2\rangle$}
  \psfrag{3}{$|3\rangle$}
  \psfrag{A1}{$A_1$}
  \psfrag{A2}{$A_2$}
  \psfrag{A3}{$A_3$}
  \psfrag{Omega3}{\hspace{-2cm} strong laser, $\Omega_3 \; \Longrightarrow$}
  \epsfig{file=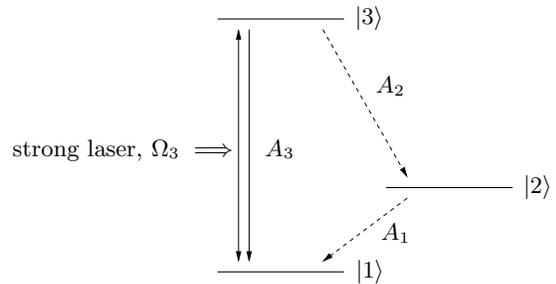,width=5cm}
  \caption{\label{Dsystem} Three-level system in D
      configuration with fast transitions (solid lines) and slow transitions
      (dashed lines).}
\end{figure}

 The experiment of Ref.~\cite{ItBeWi:88} used two
Hg$^+$ ions, with the relevant three levels as in Fig.~\ref{Dsystem}, which we
call a D configuration. 
From a theoretical point of view two such systems were
studied in Ref.~\cite{SkZaAgWeWa:01b} for the special case
$\lambda_1,~\lambda_3 \ll r \ll \lambda_2$, where $r$ is the distance
and the wave lengths
refer to the respective transitions of Fig.~\ref{Dsystem}, and for this case no
cooperative effects were found. The general case will be treated
explicitly further below. The cooperative effects
found here are of similar magnitude as for V systems, and for distances of the
above range the result of Ref.~\cite{SkZaAgWeWa:01b} is confirmed. Our
results are also in agreement with the experimentally observed absence
of cooperative effects for distances of  about 15 wave lengths
\cite{ItBeWi:88}. 

\begin{figure}[t,b]
   \psfrag{A1}{$A_1$}
  \psfrag{A2}{$A_2$}
  \psfrag{A3}{$A_3$}
  \psfrag{A4}{\hspace{-0.1cm}$A_4$}
  \psfrag{W}{$W$}
  \psfrag{Omega3}{$\Omega_3$}
  \psfrag{1}{$|1\rangle$}
  \psfrag{2}{$|2\rangle$}
  \psfrag{3}{$|3\rangle$}
  \psfrag{4}{$|4\rangle$}
  \psfrag{6P3/2}{\hspace{-0.3cm}$6\,{}^2\text{P}_{3/2}$}
  \psfrag{6P1/2}{\hspace{-0.3cm}$6\,{}^2\text{P}_{1/2}$}
  \psfrag{6S1/2}{\hspace{-0.3cm}$6\,{}^2\text{S}_{1/2}$}
  \psfrag{5D3/2}{$5\,{}^2\text{D}_{3/2}$}
  \psfrag{5D5/2}{$5\,{}^2\text{D}_{5/2}$}
  \psfrag{(a)}{$\text{(a)}$}
  \psfrag{(b)}{$\text{(b)}$}
  \epsfig{file=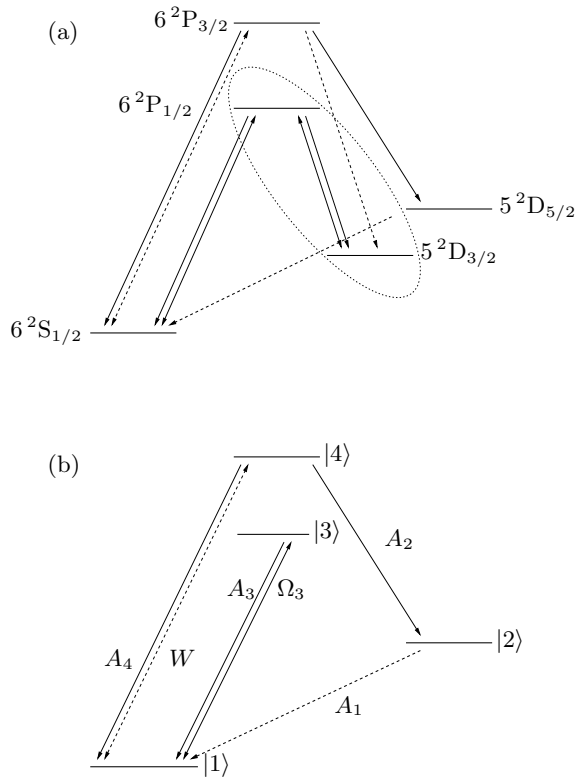, width=6.5cm, height=10cm}
    \caption{\label{5Niveau} (a) Relevant level scheme of 
    $\text{Ba}^+$ \cite{SaBlNeTo:86,Sa:86}. For circled levels see text.
(b) Effective
    four-level system for Ba$^+$. Strong coherent driving of the
    $|1\rangle - |3\rangle$ transition by a laser, weak incoherent
    driving of the $|1\rangle - |4\rangle$ transition by a lamp,
    weak decay of level $|2\rangle$.}
  \hfill
\end{figure}
The levels of Ba$^+$ used in the experiment of
Refs.~\cite{SaBlNeTo:86,SaNeBlTo:86} are 
depicted in Fig.~\ref{5Niveau} (a). The ground state
$6\,{}^2\text{S}_{1/2}$ and  the two upper states
$6\,{}^2\text{P}_{1/2}$ and $5\,{}^2\text{D}_{5/2}$ constitute a
strongly driven fluorescing $\Lambda$ system which provides the light
periods. Only when the system is in the ground state can the weak
incoherent driving of the $6\,{}^2\text{S}_{1/2}$ -
$6\,{}^2\text{P}_{3/2}$ transition populate the meta-stable
$5\,{}^2\text{D}_{5/2}$ state, with ensuing dark period. Therefore the
details of the two upper states of the $\Lambda$ system play no
significant role for the transition to a dark period, and therefore
these two states are here replaced  by  an {\em
  effective} single level. This leads to the
four-level configuration of Fig.~\ref{5Niveau} (b). The present paper
is, to our knowledge, the first to theoretically investigate possible
cooperative effects for two such four-level systems. Surprisingly,
these effects turn out to be much smaller than for two V systems for
distances $r \ge \lambda_3$, and this shows
that cooperative effects sensitively depend on how the meta-stable
level is populated. Our results for two four-level configurations are
at odds with the experimental findings of
Ref.~\cite{SaBlNeTo:86} on the magnitude of double jump rates for two
Ba$^+$ ions \cite{remark}. 

The methods presented in this paper can be carried over to
describe the Ca$^+$ experiments of
Refs.~\cite{BlReSeWe:99,DoLuBaDoStStStSt:00} although this would of
course require the use of a different level system.

The plan of the paper is as follows. In Section \ref{2Dpert} we treat
two dipole-dipole interacting D systems and explicitly calculate the
transition rates between the various light and dark periods as well as the
double jump rate. This is done by means of Bloch equations. In Section
\ref{4level} the method is carried over to two four-level systems of
Fig.~\ref{5Niveau} (b) and the transition rates are
calculated. In the Appendix a direct quantum jump approach \cite{QJ}
is outlined for two D systems.

\section{Two dipole-interacting D systems}
\label{2Dpert}

The D configuration, as displayed in Fig.~\ref{Dsystem}, is a model of
the level system of $\text{Hg}^+$ used in the 
experiments of Refs.~\cite{ItBeWi:88,ItBeHuWi:87}. The
transition $|1\rangle  
- |3\rangle$ is driven by a strong laser. Level
$|3\rangle$ can also decay via a slow transition to the meta-stable
level $|2\rangle$.  For simplicity all transitions are treated as
dipole transitions.  

\begin{figure}[t,b]
   \psfrag{g}{$|g\rangle$}
  \psfrag{s12}{$|s_{12}\rangle$}
  \psfrag{a12}{$|a_{12}\rangle$}
  \psfrag{e2}{$|e_2\rangle$}
  \psfrag{s23}{$|s_{23}\rangle$}
  \psfrag{a23}{$|a_{23}\rangle$}
  \psfrag{e3}{$|e_3\rangle$}
  \psfrag{s13}{$|s_{13}\rangle$}
  \psfrag{a13}{$|a_{13}\rangle$}
  \psfrag{(a)}{$\text{(a)}$}
  \psfrag{(b)}{$\text{(b)}$}
  \centering
  \epsfig{file=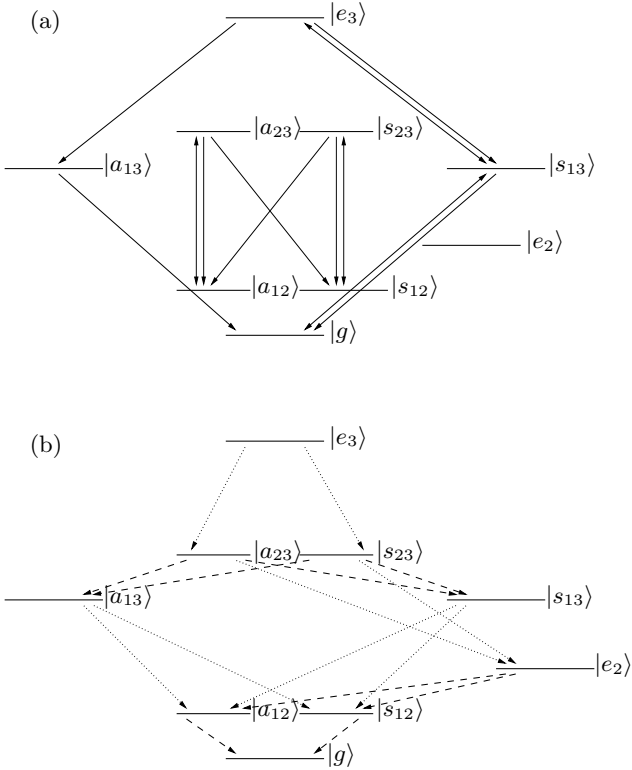, width=8cm, height=10cm}
    \caption{\label{2DNiveaus} {Level configuration of two D systems in the
      Dicke basis. (a) Slow transitions omitted. (b) Transitions with
      rate $A_2 \pm \text{Re}\,C_2$ (dotted arrows) and transitions with
      rate $A_1 \pm \text{Re}\,C_1$ (dashed arrows). Line shifts due to
      detuning and to {\rm Im}\,$C_i$ are omitted.}} 
\end{figure}

In the following we will investigate two dipole-inter\-acting D systems
a fixed distance $r$ apart and calculate the transition rates between
the three types of fluorescence periods. This will be done in this
section by means of Bloch equations. In the Appendix  the 
efficient quantum jump approach will be applied to  two D
systems, not only confirming the Bloch equation result but also giving
higher-order terms. For simplicity the
laser direction will be taken as perpendicular to the line joining the
two systems. Rabi frequency and Einstein coefficients satisfy
\begin{equation}
  A_3,\Omega_3 \gg A_1,A_2.
\end{equation}
It is convenient to use a Dicke basis,
\begin{eqnarray}
  |g\rangle & = & |1\rangle|1\rangle,\quad |e_2\rangle = |2\rangle|2\rangle, \quad
  |e_3\rangle = |3\rangle|3\rangle
\nonumber \\
  |s_{ij}\rangle & = & \big(|i\rangle|j\rangle +
  |j\rangle|i\rangle\big)/\sqrt{2}, \\
  |a_{ij}\rangle & = &
  \big(|i\rangle|j\rangle -
  |j\rangle|i\rangle\big)/\sqrt{2}\text{i}~. \nonumber 
\label{Dicke} 
\end{eqnarray}
In Fig.~\ref{2DNiveaus} (a) the level configuration for two D systems is
displayed in this basis, with the slow decays neglected, while in
Fig.~\ref{2DNiveaus} (b)  only the slow decays are shown. 
From Fig.~\ref{2DNiveaus} (a) it is
seen that without slow decays (i.e. for $A_1 = A_2 = 0$) the
configuration decouples into three independent subspaces, denoted by
$\mathcal{S}_0, \mathcal{S}_1, \mathcal{S}_2$, with 
\begin{eqnarray}\label{duals}
  \mathcal{S}_0 & = & \{|e_2\rangle\} \nonumber \\ 
  \mathcal{S}_1 & = & \{|s_{12}\rangle, |a_{12}\rangle,
    |s_{23}\rangle, |a_{23}\rangle\} \\
    \mathcal{S}_2 & = & \{|g\rangle, |s_{13}\rangle,
    |a_{13}\rangle, |e_3\rangle\}~. \nonumber 
  \end{eqnarray}

By means of the conditional Hamiltonian, $H_{\text{cond}}$, and  the reset
operation, $\mathcal{R}$, the Bloch equations can be written in the form
\cite{He:93}
\begin{equation}
\label{bloch}
  \dot{\rho}=-\frac{\text{i}}{\hbar}\left[H_{\text{cond}}\rho-\rho H_{\text{cond}}^{\dagger}\right]+ \mathcal{R}(\rho)~.
\end{equation}
 For two D systems one finds by the same method as in Refs.~\cite{AdBeDaHe:01,Bei:97}
\begin{eqnarray}
  \lefteqn{H_{\text{cond}} =} \nonumber \\ && \frac{\hbar}{2\text{i}}\Big\{A_1
  \Big[2|e_2\rangle\langle e_2| 
  + \sum_{j=1}^2 |s_{jj+1}\rangle\langle s_{jj+1}| +
  |a_{jj+1}\rangle\langle a_{jj+1}| \Big]  \nonumber \\ && \hspace{0.4cm} {}
  + \big(A_2+A_3+2\text{i}\Delta_3\big)\Big[2|e_3\rangle\langle e_3| 
  \nonumber \\ && \hspace{3.6cm} {}+
  \sum_{j=1}^2|s_{j3}\rangle\langle s_{j3}| + |a_{j3}\rangle\langle
  a_{j3}| \Big] \nonumber \\
  &&  \hspace{0.4cm} {}+ \sum_{j=1}^2 C_j\Big[|s_{jj+1}\rangle\langle
  s_{jj+1}| - |a_{jj+1}\rangle\langle a_{jj+1}|\Big] \nonumber \\ &&
  \hspace{1cm} {}
  + C_3 \Big[|s_{13}\rangle\langle
  s_{13}| - |a_{13}\rangle\langle a_{13}| \Big] \Big\} \nonumber \\
    && {}+ \frac{\hbar}{2}\Omega_3\Big\{\sqrt{2}\big(|g\rangle\langle
    s_{13}| + |s_{13}\rangle\langle e_3| \big) 
\nonumber \\ && \hspace{1.5cm} {}+ |s_{12}\rangle\langle
    s_{23}| - |a_{12}\rangle\langle a_{23}| + \mbox{h.c.} \Big\}
\end{eqnarray}
where $\Delta_3$ is the detuning of the laser. The complex dipole coupling constants $C_j$ depend in an oscillatory
way on the distance $r$,
\begin{eqnarray}
  C_j & = &
  \frac{3A_j}{2}\text{e}^{\text{i}k_jr}\left[\frac{1}{\text{i}k_jr}(1-\cos^2{\theta_j}) \right. \nonumber \\* && \left. {} +\left( \frac{1}{(k_jr)^2}-\frac{1}{\text{i}(k_jr)^3}\right)\left(1-3\cos^2{\theta_j}\right)\right]~,
\end{eqnarray}
with $k_j = 2\pi/\lambda_j$ and $\theta_j$  the angle between the
corresponding dipole 
moment and the line connecting the systems. For maximal effect we take
$\theta_j= \pi/2$ in the following. The real part of $C_j$ leads to  changes of the decay
constants and  the imaginary part to a level shift in the Dicke basis, as
seen from the expression for $H_{\rm cond}$.
The reset operation can be written in the form
\begin{eqnarray}
  \label{2DRucksetz}
  \mathcal{R} (\rho)  & = & \sum_{j=1}^3 \left[(A_j + \text{Re} C_j)R_+^{(j)}
    \rho \hat{R}_+^{(j)^{\scriptstyle \dagger}} 
    \nonumber \right. \\* && \left. \hspace{1cm} {}+
    (A_j-\text{Re} C_j) R_-^{(j)} \rho
    R_-^{(j)^{\scriptstyle \dagger}} \right] ,
\end{eqnarray}
where
\begin{eqnarray*}
  R_+^{(1)} & = & |g\rangle\langle s_{12}|
  + |s_{12}\rangle\langle e_2| 
  \\* && {}
  + \frac{1}{\sqrt{2}}\big(|s_{13}\rangle\langle s_{23}| 
  + |a_{13}\rangle\langle a_{23}| \big) \\ 
  R_-^{(1)} & = & |g\rangle\langle a_{12}|
  + |a_{12}\rangle\langle e_2| 
  \\* && {}
  + \frac{1}{\sqrt{2}}\big(|a_{13}\rangle\langle s_{23}| -
  |s_{13}\rangle\langle a_{23}| \big) \\  
  R_+^{(2)} & = & |e_2\rangle\langle s_{23}| + |s_{23}\rangle\langle e_3| 
  \\* && {}
  + \frac{1}{\sqrt{2}}\big(|s_{12}\rangle\langle s_{13}| 
  + |a_{12}\rangle\langle a_{13}| \big) \\ 
  R_-^{(2)} & = & |e_2\rangle\langle
  a_{23}| + |a_{23}\rangle\langle e_3| 
  \\* && {}
  + \frac{1}{\sqrt{2}}\big(|s_{12}\rangle\langle a_{13}| 
  - |a_{12}\rangle\langle s_{13}| \big) \\ 
  R_+^{(3)} & = & |g\rangle\langle s_{13}|
  + |s_{13}\rangle\langle e_3| 
  \\* && {}
  + \frac{1}{\sqrt{2}}\big(|s_{12}\rangle\langle s_{23}| 
  - |a_{12}\rangle\langle a_{23}|\big) \\
  R_-^{(3)} & = & |g\rangle\langle a_{13}|
  + |a_{13}\rangle\langle e_3| 
  \\* && {}
  + \frac{1}{\sqrt{2}}\big(|s_{12}\rangle\langle a_{23}| 
  + |a_{12}\rangle\langle s_{23}|\big)~.   
\end{eqnarray*}
To determine the transition rates we write Eq.~(\ref{bloch}) in a
Liouvillean form as 
\begin{equation}
  \dot{\rho}=\mathcal{L}\rho = \left\{ \mathcal{L}_0 (A_3, C_3,
    \Omega_3,\Delta_3)  + \mathcal{L}_1 (A_1, A_2, C_1, C_2 ) \right\} \rho
\end{equation}
where the super-operator $\mathcal{L}_1(A_1, A_2, C_1, C_2)$ is a
perturbation depending on the small parameters, and employ the
following important property of the
time development. Starting with an initial state in one of the
subspaces $\mathcal{S}_i$, the system will rapidly---on a time scale
of $\Omega_3^{-1}$ and $A_3^{-1}$---approach one of the
quasi-stationary states $\rho_{\text{ss},i}$. Thereafter---for times much
larger than $\Omega_3^{-1}$ and $A_3^{-1}$, but much smaller than
$A_1^{-1}$ and $A_2^{-1}$---small populations in the other subspaces
will build up until eventually,
on a time scale of $A_1^{-1}$ and $A_2^{-1}$, the true stationary
state is approached. Hence we consider a time $\Delta t$ with
\begin{equation}
\label{relA3A1}
  A_3^{-1},\Omega_3^{-1} \ll \Delta t \ll A_1^{-1},A_2^{-1}
\end{equation}
and calculate $\rho(t_0+\Delta t)$ for initial
$\rho(t_0)=\rho_{\text{ss},i}$. The quasi-invariant states are easily
calculated from $\mbox{$\mathcal{L}_0\rho_{\text{ss},i}=0$}$ as 
\begin{eqnarray}
  \label{rho0dark} 
  \lefteqn{
  \rho_{\text{ss},0} = |e_2\rangle\langle e_2|,} \hspace{1.5cm} \\
  \label{rho0inner}
  \lefteqn{
    \rho_{\text{ss},1} = 
  \frac{1}{2}\frac{A_3^2+\Omega_3^2+4\Delta_3^2}{A_3^2+2\Omega_3^2+4\Delta_3^2} 
  \big( |s_{12}\rangle\langle s_{12}| + |a_{12}\rangle\langle a_{12}| \big)}
  \hspace{1.5cm} \\ && \hspace{-1.5cm} 
  {}+ \frac{1}{2}\frac{\Omega_3^2}{A_3^2+2\Omega_3^2+4\Delta_3^2}
  \big( |s_{23}\rangle\langle s_{23}| + |a_{23}\rangle\langle a_{23}|
  \big) \nonumber \\ && \hspace{-2.5cm} 
  {}+ \left\{\frac{1}{2}\frac{(\text{i}A_3-2\Delta_3)
      \Omega_3}{A_3^2+2\Omega_3^2+4\Delta_3^2}
  \big( |s_{12}\rangle\langle s_{23}| - |a_{12}\rangle\langle a_{23}|\big)
  + \mbox{h.c.} \right\} \nonumber \\
  \label{rho0outer}
  \lefteqn{
    \rho_{\text{ss},2} = } \hspace{1.5cm} \\ && \hspace{-2.5cm}
  \frac{1}{N}\Big[\big\{N - \Omega_3^2(2A_3^2+3\Omega_3^2+8\Delta_3^2)
  \big\}|g\rangle\langle g| 
  \nonumber \\ && \hspace{-2cm} {}
  + \Omega_3^2(2A_3^2+\Omega_3^2+8\Delta_3^2)|s_{13}
  \rangle\langle s_{13}| \nonumber \\ && \hspace{-2cm} {}
  + \Omega_3^4\big\{|e_3\rangle\langle e_3|
  +|a_{13}\rangle\langle a_{13}| \big\}
  \nonumber \\ && \hspace{-2cm} 
  {}+ \Big\{\Omega_3(\text{i}A_3-2\Delta_3)\Big(
  \sqrt{2}(A_3^2+\Omega_3^2+4\Delta_3^2 \nonumber \\ && \hspace{1.5cm} {}
  + (A_3-2\text{i}\Delta_3)C_3)|g \rangle \langle s_{13}| \nonumber \\
  && \hspace{1cm} 
  {}+ \Omega_3(\text{i}A_3-2\Delta_3+\text{i}C_3)
  |g\rangle\langle e_3| \nonumber \\ && \hspace{1cm} {}
  + \sqrt{2}\Omega_3^2 |s_{13}\rangle\langle e_3|\Big) +
  \mbox{h.c.}\Big\} \Big] \nonumber \\
  \mbox{where} \qquad 
  N & = & (A_3^2+2\Omega_3^2+4\Delta_3^2)^2 \nonumber \\ && \hspace{-1.5cm} 
  {}+ (A_3^2+4\Delta_3^2)(|C_3|^2 +
  2A_3\text{Re}\,C_3 + 4\Delta_3\text{Im}\,C_3) \nonumber
\end{eqnarray}
As in Ref.~\cite{AdBeDaHe:01} one has in
perturbation theory
\begin{eqnarray}
  \label{rhoDeltatA1}
  \lefteqn{
 \rho(t_0+\Delta t) = } 
 \\ && 
  \rho_{\text{ss},i} + \int\limits_0^{\Delta
    t}\text{d}\tau \,
  \text{e}^{\mathcal{L}_0t} \mathcal{L}_1\rho_{\text{ss},i} +
  \mathcal{O}\big((A_1,A_2,C_1,C_2)^2\big), \nonumber
\end{eqnarray}
but, unlike Ref.~\cite{AdBeDaHe:01}, $\mathcal{L}_1\rho_{\text{ss},i}$ 
is not a superposition of just the eigenstates for nonzero eigenvalues
of $\mathcal{L}_0$ but also of the $\rho_{\text{ss},j}$'s. We therefore decompose
$\mathcal{L}_1\rho_{\text{ss},i}$  into a superposition of all
eigenstates (matrices) of $\mathcal{L}_0$, 
\begin{equation}
  \label{L1rho0}
  \mathcal{L}_1\rho_{\text{ss},i}=\sum_{j=0}^2 \alpha_{ij}\rho_{\text{ss},j} + \tilde{\rho}
\end{equation}
where $\tilde{\rho}$ contains the contributions from the eigenstates
for nonzero eigenvalues of $\mathcal{L}_0$. For later use we note that
these eigenvalues are of the order of $A_3$ and $\Omega_3$. The coefficients
$\alpha_{ij}$ can easily be determined by means of the reciprocal (or
dual) eigenstates where only those for eigenvalue $0$ of
$\mathcal{L}_0$ are needed. They are denoted by $\rho_{\text{ss}}^i$ and
are defined through
\begin{equation}
\label{norm}
\text{Tr}(\rho_{\text{ss}}^{i\dagger}\rho_{\text{ss},j})=\delta_{ij}, \qquad i,j = 0,1,2
\end{equation}
\begin{equation}
\label{dual}
\text{Tr}(\rho_{\text{ss}}^{i\dagger}\mathcal{L}_0A)=0 \quad \mbox{for any matrix $A$.}
\end{equation}
The latter means $\mathcal{L}_0^{\dagger}\rho^i_{\text{ss}}=0$, with
the adjoint 
$\mathcal{L}_0^{\dagger}$ defined with respect to a scalar product
given by 
$\text{Tr}(A^{\dagger}B)$. Then one has
\begin{equation}
\label{alpha}
  \alpha_{ij}=\text{Tr}(\rho_{\text{ss}}^{j\dagger}\mathcal{L}_1\rho_{\text{ss},i})~.
\end{equation}
The reciprocals $\rho_{\text{ss}}^i$ are easily determined as
follows. Since the Bloch equations conserve the trace one has
\begin{equation}
  0=\text{Tr}\dot{\rho}=\text{Tr}\mathcal{L}_0\rho
\end{equation}
for any $\rho$. Thus
\begin{equation}
\label{dualone}
  0=\text{Tr}(\openone \mathcal{L}_0\rho)=\text{Tr} ((\mathcal{L}_0^\dagger \openone) \rho)
\end{equation}
for any $\rho$ and therefore $\mathcal{L}_0^{\dagger}\openone=0$. Now
$\openone$ can be written as a sum of terms purely from $\mathcal{S}_0,
\mathcal{S}_1$, and $\mathcal{S}_2$ and, since the subspaces are
invariant under $\mathcal{L}_0$, these terms must be annihilated by
$\mathcal{L}_0^{\dagger}$ individually. This yields
\begin{eqnarray*}
  \rho^0_{\text{ss}} & = & |e_2\rangle\langle e_2|, \\
  \rho^1_{\text{ss}} & = & |s_{12}\rangle\langle s_{12}| + |a_{12}\rangle\langle
  a_{12}| 
  + |s_{23}\rangle\langle s_{23}| + |a_{23}\rangle\langle a_{23}|, \\ 
  \rho^2_{\text{ss}} & = & |g\rangle\langle g| +
  |s_{13}\rangle\langle s_{13}| 
  + |a_{13}\rangle\langle a_{13}| + |e_3\rangle\langle e_3| 
\end{eqnarray*}
since the sum of the right-hand sides indeed yields $\openone$ and the
normalization condition of Eq.~(\ref{norm}) is fulfilled. From
Eq.~(\ref{alpha}) one then obtains the $\alpha_{ij}$.
Inserting now $\mathcal{L}_1\rho_{\text{ss},i}$ into
Eq.~(\ref{rhoDeltatA1}) one obtains
\begin{subequations}
\begin{eqnarray}
  \label{rhotdeltat}
  \lefteqn{\rho(t_0+\Delta t) = } \nonumber \\ &&
  \rho_{\text{ss},i} + \int\limits_0^{\Delta
    t}\text{d}\tau \,\left(\sum_{j=0}^2\alpha_{ij}\rho_{\text{ss},j} +
    \text{e}^{\mathcal{L}_0\tau}\tilde{\rho}\right) \\ \label{rhotdeltata}
  & = & \rho_{\text{ss},i} + \sum_{j=0}^2
  \alpha_{ij}\rho_{\text{ss},j}\Delta t + (\epsilon-\mathcal{L}_0)^{-1}\tilde{\rho} 
\end{eqnarray}
\end{subequations} 
where for the $\tilde{\rho}$ term the upper integration limit can be
extended to infinity since $\tilde{\rho}$ belongs to nonzero
eigenvalues of $\mathcal{L}_0$ and is therefore rapidly damped. Now,
$\mathcal{L}_0^{-1}$ is of the order of $A_3^{-1}$ and $\Omega_3^{-1}$
on $\tilde{\rho}$, and thus the last term in Eq.~(\ref{rhotdeltata})
is of the order of $\mathcal{L}_1/(A_3,\Omega_3)$ which is much
smaller than $\alpha_{ij}\Delta t \sim \mathcal{L}_1\Delta t$, by
Eq.~(\ref{relA3A1}). Therefore the last term in Eq.~(\ref{rhotdeltata})
can be neglected, and this equation then reveals that the $\alpha_{ij}$'s
have the meaning of transition rates from subspace $\mathcal{S}_i$ to
$\mathcal{S}_j$, i.e.,
\begin{equation}
\label{pij}
p_{ij}=\alpha_{ij}.
\end{equation} 
 
The transition rates are now obtained from
Eqs.~(\ref{pij}), (\ref{alpha}), and (\ref{rho0dark})-(\ref{rho0outer}) as
\begin{subequations}
\begin{eqnarray}
  p_{01} & = & 2A_1 \\
  p_{10} & = & \frac{A_2\Omega_3^2}{A_3^2+2\Omega_3^2+4\Delta_3^2} \\ 
  p_{12} & = & A_1 
\end{eqnarray}
and
\begin{widetext}
\begin{eqnarray}
  \label{2Dp21}
  p_{21} & = & 
  2\frac{A_2\Omega_3^2[A_3^2+2\Omega_3^2+4\Delta_3^2]}{[A_3^2 +
    2\Omega_3^2 + 4\Delta_3^2]^2 + [A^2_3+4\Delta^2_3][|C_3|^2+
    2A_3\text{Re}\,C_3+ 4\Delta_3\text{Im}\,C_3]} \nonumber \\
  & = & \frac{2A_2\Omega_3^2}{A_3^2+2\Omega_3^2+4\Delta_3^2}\left[1
    - 2\text{Re}\,C_3\frac{A_3(A_3^2+4\Delta_3^2)}{(A_3^2 +
      2\Omega_3^2 + 4\Delta_3^2)^2} -
    4\text{Im}\,C_3\frac{\Delta_3(A_3^2 + 4\Delta_3^2)}{(A_3^2 +
      2\Omega_3^2+4\Delta_3^2)^2} \right] + \mathcal{O}(C_3^2)~.
\end{eqnarray}
\end{widetext}
\end{subequations}
One sees that, for two D systems, only $p_{21}$ depends on the dipole
coupling constants $C_3$ in first order. This contrasts with two V
systems where both $p_{21}$ and $p_{12}$ depend on $C_3$. The physical
reason for this is that transitions between bright periods for two D
systems are due to decays and not due to the laser. Hence the transition
rates should essentially be governed by the Einstein coefficients of these
decays on the one hand and by the population of the states in the
initial subsystem on the other hand. Therefore the parameter $C_3$
enters only through the quasi-stationary state $\rho_{\text{ss},i}$ of the
initial subsystem. The absence of a linear
$C_1$ and $C_2$ dependence can be understood from
Fig.~{\ref{2DNiveaus}} (b) as
follows. For most slow transitions between two subspaces there are two
channels with rates $A_j \pm \text{Re}\, C_j$ so that $\text{Re}\, C_j$
cancels. States with a single decay channel lie in $\mathcal{S}_1$ and, by
symmetry, they appear in pairs with different sign of $\text{Re}\, C_j$.

\begin{figure}[t!]
  \psfrag{p21}{$p_{21} [s^{-1}]$}
  \psfrag{nDJ}{$n_{\text{DJ}} [10^{-6}s^{-1}]$}
  \psfrag{r}{\hspace{-0.4cm}$r/\lambda_3$}
   \epsfig{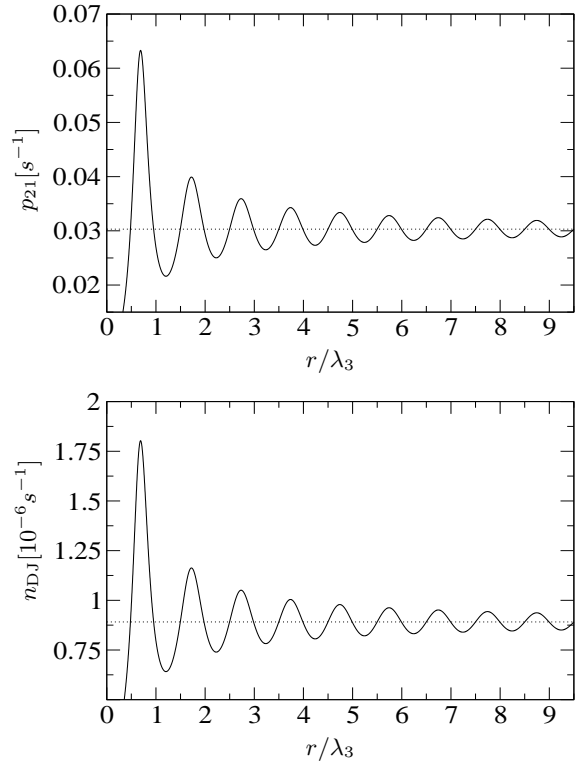}
   \caption{\label{2Dp21nDS}  Transition rate $p_{21}$ and
     double jump rate $n_{\text{DJ}}$ for two dipole-interacting D
     systems. The dashed lines show the case of independent
     systems. Parameter values are $A_1=1\,s^{-1}$, $A_2=1\,s^{-1}$,
     $A_3=4\cdot 10^8 \,s^{-1}$, $\Omega_3=5\cdot10^7\,s^{-1}$, and
     $\Delta_3=0$.} 
\end{figure}

From the transition rates $p_{ij}$ one obtains the double jump rate,
$n_{\text{DJ}}$, by the formula \cite{AdBeDaHe:01}
\begin{equation}
  n_{\text{DJ}}=2
  \frac{p_{10}p_{01}p_{12}p_{21}} {p_{01}p_{21} + p_{21}p_{10} +
    p_{01}p_{12}} \Delta T_{\text{DJ}}
\end{equation}
where $\Delta T_{\text{DJ}}$ is the defining small time interval for a double
jump. Significant cooperative effects occur only as
long as $\Omega_3$ and $\Delta_3$ are at least an order of magnitude
smaller than $A_3$. When compared to non interacting systems, the
cooperative effects are up to 30\% for distances between one and two
wave lengths and 5\% around ten wave wave lengths, similar as for two
V systems. For longer distances they are practically absent and this
is consistent with the experimental results of Ref.~\cite{ItBeWi:88}.
Fig.~\ref{2Dp21nDS} shows $p_{21}$ and $n_{\text{DJ}}$ 
versus the relative distance $r/\lambda_3$ for
typical parameters, where $\lambda_3$ is the wave length of the strong
transition. 

Our explicit results for arbitrary $r$ confirm the large-distance
result of Ref.~\cite{SkZaAgWeWa:01b}   who argued that for $\lambda_1,\lambda_3 \ll r \ll \lambda_2$ cooperative effects are ``suppressed by the rapid decay on
the fast transition''. In fact, we find in the Appendix that these
effects are only to first order independent of the coupling parameter
$C_2$; the second order contributions in $C_2$ are, however,
negligible for the experimental values of Ref.~\cite{ItBeWi:88}.

\section{Two dipole-interacting four-level systems as a model for two
  B\lowercase{a}$^+$ ions}
\label{4level}
The experiments of Refs.~\cite{Sa:86,SaBlNeTo:86} used
$\text{Ba}^+$. As explained in the introduction we model the relevant
level scheme  by the effective four-level configuration in
Fig.~\ref{5Niveau} (b).  
The $|1\rangle - |4\rangle$ transition is driven weakly
and incoherently by a lamp, while the $|1\rangle - |3\rangle$ transition is driven coherently by a strong laser. This
time the Bloch equation can be written as \cite{HePl:93}
\begin{eqnarray}
  \label{Bloch4N}
  \dot{\rho} & = & -\frac{\text{i}}{\hbar} \Big[ H_{\text{cond}} \rho - \rho
  H_{\text{cond}}^{\dagger}\Big] +
  \mathcal{R}_W(\rho) + \mathcal{R}(\rho)\\
  & \equiv & \left\{\mathcal{L}_0^{\dagger(0)}(A_2,A_3,A_4,\Omega_3,\Delta_3,C_3)
+ \mathcal{L}_0^{\dagger(1)}(C_2,C_4)\right\}\rho \nonumber
\end{eqnarray}
where $\mathcal{R}_{\text{W}}(\rho)$ describes the incoherent driving as
in Ref.~\cite{HePl:93} and is given explicitly below. The Dicke states are
defined in analogy to Eq.~(\ref{Dicke}), and $H_{\text{cond}}$ and
$\mathcal{R}(\rho)$ can be calculated as in
Refs.~\cite{AdBeDaHe:01,Bei:97} as
\begin{widetext}
\begin{eqnarray}
H_{\text{cond}} & = & \frac{\hbar}{2\text{i}}\Big\{ A_1\Big[2|e_2\rangle\langle e_2| +
    |s_{12}\rangle\langle s_{12}| + |a_{12}\rangle\langle a_{12}| +
    |s_{23}\rangle\langle s_{23}| + |a_{23}\rangle\langle a_{23}| +
    |s_{24}\rangle\langle s_{24}| + |a_{24}\rangle\langle a_{24}|\Big] 
    \nonumber \\ && \hspace{0.5cm} 
    {}+ \big(A_2 + A_4\big)\Big[2|e_4\rangle\langle
    e_4| + \sum_{j=1}^3 \left\{|s_{j4}\rangle\langle s_{j4}| +
    |a_{j4}\rangle\langle a_{j4}|\right\} \Big] 
  \nonumber \\ && \hspace{0.5cm} {}
    + \big(A_3+2\text{i}\Delta_3\big)\Big[2|e_3\rangle\langle e_3| +
    |s_{13}\rangle\langle s_{13}| +
    |a_{13}\rangle\langle a_{13}| + |s_{23}\rangle\langle s_{23}| +
    |a_{23}\rangle\langle a_{23}| + |s_{34}\rangle\langle s_{34}| +
    |a_{34}\rangle\langle a_{34}| 
    \Big] \nonumber \\
    && \hspace{0.5cm}{}+ W\Big[2|g\rangle\langle g| +
    2|e_4\rangle\langle e_4| +
    \sum_{j=1}^3 \left\{|s_{j4}\rangle\langle s_{j4}| +
    |a_{j4}\rangle\langle a_{j4}| \right\} + \sum_{j=2}^4
  \left\{|s_{1j}\rangle\langle s_{1j}| + |a_{1j}\rangle\langle a_{1j}|\right\}
    \Big] \Big\} \nonumber \\
    && \hspace{-0.5cm} {}+ \frac{\hbar}{2\text{i}}\Big\{
    C_2\big(|s_{23}\rangle\langle s_{23}| - |a_{23}\rangle\langle
    a_{23}|\big) + 
    C_3\big(|s_{13}\rangle\langle s_{13}| - |a_{13}\rangle\langle
    a_{13}|\big) + C_4\big(|s_{14}\rangle\langle s_{14}| -
    |a_{14}\rangle\langle a_{14}|\big) \Big\}
    \nonumber \\
    && \hspace{-0.5cm} {}+ \frac{\hbar}{2}\Omega_3\Big\{\sqrt{2}\big(|g\rangle\langle
    s_{13}| + |s_{13}\rangle\langle e_3| \big) + |s_{12}\rangle\langle
    s_{23}| - |a_{12}\rangle\langle a_{23}| + |s_{14}\rangle\langle
    s_{34}| + |a_{14}\rangle\langle a_{34}| + \mbox{h.c.} \Big\}
\end{eqnarray}
\end{widetext}
\begin{eqnarray}
  \mathcal{R}(\rho) & = & 
  \sum_{j=1}^4
  \left[\big(A_j+\text{Re}\,C_j\big)R_{+}^{(j)}\rho
    R_{+}^{(j)^{\scriptstyle \dagger}} 
  \right. \nonumber \\*  && \left. \hspace{0.6cm} {}
    + \big(A_j-\text{Re}\,C_j\big)R_{-}^{(j)}\rho
    R_{-}^{(j)^{\scriptstyle \dagger}}\right]  
\end{eqnarray}
with
\begin{eqnarray*}
  R_+^{(1)}& = & |g\rangle\langle
  s_{12}|+|s_{12}\rangle\langle  e_2|  \\* && \hspace{-1cm} {}
  + \frac{1}{\sqrt{2}} \big(|s_{13}\rangle\langle
  s_{23}| +  |a_{13}\rangle\langle a_{23}| + |s_{14}\rangle\langle
  s_{24}| +  |a_{14}\rangle\langle a_{24}| \big) \\
  R_-^{(1)} & = & |g\rangle\langle
  a_{12}|+|a_{12}\rangle\langle  e_2| \\* && \hspace{-1cm} {}
  +\frac{1}{\sqrt{2}} \big(|a_{13}\rangle\langle
  s_{23}| -  |s_{13}\rangle\langle a_{23}| + |a_{14}\rangle\langle
  s_{24}| -  |s_{14}\rangle\langle a_{24}| \big) \\
  R_+^{(2)} & = & |e_2\rangle\langle
  s_{24}|+|s_{24}\rangle\langle e_4| \\* && \hspace{-1cm} {}
  + \frac{1}{\sqrt{2}}\big(|s_{12}\rangle\langle s_{14}| +
  |a_{12}\rangle\langle a_{14}| + |s_{23}\rangle\langle s_{34}| -
  |a_{23}\rangle\langle a_{34}|\big) \\
  R_-^{(2)} & = & |e_2\rangle\langle
  a_{24}|+|a_{24}\rangle\langle e_4| \\* && \hspace{-1cm} {}
  + \frac{1}{\sqrt{2}}\big(|s_{12}\rangle\langle a_{14}| -
  |a_{12}\rangle\langle s_{14}| + |s_{23}\rangle\langle a_{34}| +
  |a_{23}\rangle\langle s_{34}|\big) \\
  R_+^{(3)} & = & |g\rangle\langle
  s_{13}|+|s_{13}\rangle\langle e_3| \\* && \hspace{-1cm} {}
  +\frac{1}{\sqrt{2}}\big(|s_{12}\rangle\langle s_{23}| -
  |a_{12}\rangle\langle a_{23}| + |s_{14}\rangle\langle s_{34}| +
  |a_{14}\rangle\langle a_{34}|\big) \\
  R_-^{(3)} & = & |g\rangle\langle a_{13}|+|a_{13}\rangle\langle e_3|
  \\* && \hspace{-1cm} {}
  + \frac{1}{\sqrt{2}}\big(|s_{12}\rangle\langle a_{23}| +
  |a_{12}\rangle\langle s_{23}| + |s_{14}\rangle\langle a_{34}| -
  |a_{14}\rangle\langle s_{34}|\big) \\
  R_+^{(4)} & = & |g\rangle\langle s_{14}|+|s_{14}\rangle\langle e_4|
  \\* && \hspace{-1cm} {}
  + \frac{1}{\sqrt{2}} \big(|s_{12}\rangle\langle
  s_{24}| - |a_{12}\rangle\langle a_{24}| + |s_{3j}\rangle\langle
  s_{34}| - |a_{13}\rangle\langle a_{34}| \big) \\ 
  R_-^{(4)} & = & |g\rangle\langle a_{14}|+|a_{14}\rangle\langle e_4|
  \\* && \hspace{-1cm} {}
  + \frac{1}{\sqrt{2}} \big(|s_{12}\rangle\langle a_{24}| + 
  |a_{2j}\rangle\langle s_{24}| + |s_{3j}\rangle\langle a_{34}| + 
  |a_{3j}\rangle\langle s_{34}| \big) \nonumber
\end{eqnarray*}
The lamp term is obtained as in Ref.~\cite{HePl:93} as
\begin{eqnarray}
  \lefteqn{\mathcal{R}_W(\rho)=} \\* && W\big(R_{4+}\rho R_{4+}^{\dagger} + R_{4-}\rho
  R_{4-}^{\dagger} + R_{4+}^{\dagger}\rho R_{4+} + R_{4-}^{\dagger}\rho
  R_{4-}\big) \nonumber
\end{eqnarray}
where $W$  is the product of the spectral energy density of the lamp
and the Einstein $B$ coefficient of the $|1\rangle -|4\rangle $
transition.

Now the procedure is similar as for the D system. The Liouvillean
$\mathcal{L}_0$ possesses three (quasi-) stationary states
$\rho_{\text{ss},0}$, $\rho_{\text{ss},1}$, and
$\rho_{\text{ss},2}$ which 
coincide with those for the D systems in
Eqs.~(\ref{rho0dark}-\ref{rho0outer}) and which are associated with the
dark and the two bright periods. As before, one calculates
$\rho(t_0+\Delta t)$ as in 
Eq.~(\ref{rhoDeltatA1}) and  decomposes
$\mathcal{L}_1\rho_{\text{ss},i}$ as in Eq.~(\ref{L1rho0}). Now, however,
the reciprocals $\rho_{\text{ss}}^i$ are more difficult to determine
since $|4\rangle$ can decay into $|1\rangle$ as well as $|2\rangle$.
 An exact solution of $\mathcal{L}_0^{\dagger}\rho_{\text{ss}}^i=0$ is rather
elaborate. We therefore decompose
\begin{equation}
\mathcal{L}_0^{\dagger}=\mathcal{L}_0^{\dagger(0)}(A_2,A_3,A_4,\Omega_3,\Delta_3,C_3)
+ \mathcal{L}_0^{\dagger(1)}(C_2,C_4).
\end{equation}
and, by Maple, have calculated  $\rho^i_{\text{ss}}$ to first order in
perturbation theory with respect to $C_2$ and $C_4$, with the same
constraint as in Eq.~(\ref{norm}). The lengthy result will not be
given here explicitly. The transition rates are again given by
\begin{equation}
p_{ij}=\text{Tr}(\rho_{\text{ss}}^{j\dagger}\mathcal{L}_1\rho_{\text{ss},i})
\end{equation} 
and one obtains for two dipole-interacting four-level systems of
Fig.~\ref{5Niveau} (b)  to first order in $C_2$ and $C_4$
\begin{subequations}
\begin{eqnarray}
  p_{01} & = & 2A_1 \\
  p_{10} & = & \frac{A_2W(A_3^2+\Omega_3^2+ 4\Delta_3^2)}{(A_2+A_4)[A_3^2 +
    2\Omega_3^2 + 4\Delta_3^2]} \\
  p_{12} & = & A_1 
\end{eqnarray}
and 
\begin{widetext}
\begin{eqnarray}
  p_{21} & = & 2A_2W\frac{(A_3^2+\Omega_3^2+4\Delta_3^2)(A_3^2 +
    2\Omega_3^2 + 4\Delta_3^2)+(A_3^2+4\Delta_3^2)(|C_3|^2+2A_3\text{Re}\, C_3 + 4\Delta_3\text{Im}\,C_3)}
  {(A_2+A_4)(A_3^2+2\Omega_3^2+4\Delta_3^2)^2+(A_3^2+4\Delta_3^2)(|C_3|^2+2A_3\text{Re}\, C_3 + 4\Delta_3\text{Im}\,C_3)} \nonumber \\
  & = & 2A_2W\Bigg[\frac{A_3^2+\Omega_3^2+
    4\Delta_3^2}{(A_2+A_4)[A_3^2 + 2\Omega_3^2 + 4\Delta_3^2]} +
  2\,\text{Re}\,C_3\frac{A_3\Omega_3^2(A_3^2+4\Delta_3^2)}{(A_2+A_4)[A_3^2 +
    2\Omega_3^2 + 4\Delta_3^2]^3} \nonumber \\
  && \hspace{5.5cm} {}+ 4\,\text{Im}\,C_3\frac{\Delta_3\Omega_3^2
    (A_3^2 +4\Delta_3^2)}{(A_2+A_4)[A_3^2 +
    2\Omega_3^2 + 4\Delta_3^2]^3}\Bigg] + \mathcal{O}(C_3^2).
\end{eqnarray}
\end{widetext}
\end{subequations}
It is seen that $p_{01}$, $p_{10}$, and $p_{12}$ are independent of
the coupling parameters and are thus the same as for non interacting
systems. 

\begin{figure}[t,b]
  \psfrag{nDJ}{$ n_{\text{DJ}} [10^{-4}s^{-1}]$}
  \psfrag{p21}{$ p_{21} [10^{-1}s^{-1}]$}
  \psfrag{r}{\hspace{-0.4cm}$r/\lambda_3$}
     \epsfig{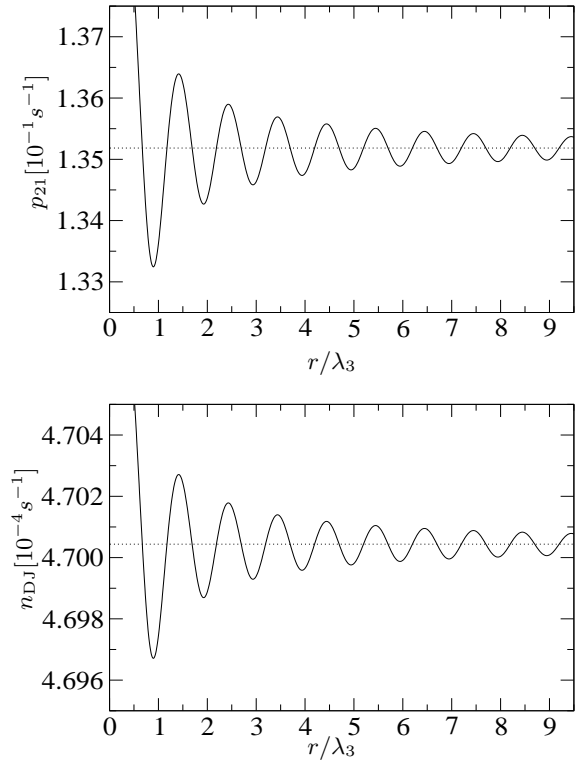} 
  \caption{\label{24nivp21nDS} Transition rate $p_{21}$ and
    double-jump rate $n_{\text{DJ}}$
    for dipole-interacting four-level systems, with optimal
    $\Omega_3=\frac{1}{2}\sqrt{\sqrt{5}-1}\sqrt{A_3^2+4\Delta_3^2}$ and all
    other parameters as in the experiment \cite{SaBlNeTo:86}. The
    dashed lines show the case of independent systems.
    } 
\end{figure}

These results for two four-level systems show great similarity with
those for the two D systems of the proceeding section. In both cases only
$p_{21}$ depends to first order on $C_3$, the coupling parameter
associated with the laser-driven transition. However, cooperative
effects are significantly smaller for the two four-level systems. For
fixed laser detuning, the effect of $C_3$ becomes maximal for $\Omega_3=\frac{1}{2}
\sqrt{\sqrt{5}-1}\sqrt{A_3^2+4\Delta_3^2}$. For this value of
$\Omega_3$, Fig.~\ref{24nivp21nDS} shows the transition rate $p_{21}$ from a
double intensity period to a unit-intensity period and the double jump
rate $n_{\text{DJ}}$ over the relative distance $r/\lambda_3$, with
the other parameters as in the experiment
\cite{SaBlNeTo:86,Sa:86}. Despite the optimal choice of the
Rabi frequency, $\Omega_3$, the deviations from the value for
non-interacting systems are very small. Already for a distance of
about a wave length $\lambda_3$, they are not more than $1\%$ for $p_{21}$
when compared to non interacting systems, while for $n_{\text{DJ}}$ they are
less than 1\textperthousand.

\section{Conclusions}
We have investigated the effect of the dipole-dipole interaction for
two fluorescing systems with macroscopic light and dark periods, first
for three-level  D configurations and then for four-level
systems. The three-level  D configuration models the relevant levels
of  Hg$^+$ used in the experiments of
Ref.~\cite{ItBeHuWi:87,ItBeWi:88}, and the four-level configuration is
an effective model for Ba$^+$,  
 used in the experiments of Ref.~\cite{SaBlNeTo:86,SaNeBlTo:86}. For
these systems one has macroscopic light and dark periods, and their
statistics can be a sensitive test of the dipole-dipole
interaction. We have explicitly calculated the transition rates
between the different light and dark periods by employing Bloch
equations as well as a direct quantum jump approach. From the
transition rates the double jump rates are obtained.

For two D systems the effect of the dipole-dipole interaction is of
similar magnitude as for two V system investigated earlier
\cite{AdBeDaHe:01} and shown to be up to 30 \% for distances of the order of a
wave length of the strong transition and about 5\% around ten wave
lengths, when compared to independent systems.  For longer distances
they are practically absent and this is in agreement with the
experimental results of Ref.~\cite{ItBeWi:88}. We have also recovered 
the special case of Ref.~\cite{SkZaAgWeWa:01b} where distances 
satisfying $\lambda_1,~\lambda_3 \ll r \ll \lambda_2$ were considered
and where  an argument for the non dependence on the dipole
coupling constant $C_2$ was given. Here we have shown that this holds
to first order and that the explicitly determined second order terms
are negligibly small.

For the effective model of two Ba$^+$ systems our
results yield very small and hardly observable cooperative effects for
the double jump rate. This is at odds with experimental result in
Ref.~\cite{SaBlNeTo:86}. Our method also applies to three Ba$^+$ ions,
but this is more tedious and requires another paper. Also a
theoretical investigation of the experiments with Ca$^+$
\cite{BlReSeWe:99,DoLuBaDoStStStSt:00} is possible with our
method. For this the calculations have to be carried over to a level
scheme modeling that of Ca$^+$.

A further conclusion of our work is the observation that the magnitude
of cooperative effects due to the dipole-dipole interaction sensitively
depends on how the meta-stable level is populated.

\appendix*

\section{Quantum jump approach for two D systems}
\label{quantjump} 
The procedure will first be explained for a single $D$ system which
has just two types of periods, light and dark ones. From its level
configuration in Fig.~\ref{Dsystem} it is evident that the onset of a
dark period is preceded by a photon from the $|3\rangle - |2\rangle$
transition, with frequency $\omega_2$. Hence, starting at $t_0=0$ in
$|1\rangle$, the probability density for the next photon to occur at
time $t$ and to come from the $|3\rangle - |2\rangle$ transition is
\begin{equation}
\label{w1omega2}
w_{1\omega_2}(t) = A_2 |\langle3| \text{e}^{-\text{i}H_{\rm cond}t/\hbar} |1\rangle|^2
\end{equation}
since $H_{\rm cond}$ gives the time development between photon
emissions \cite{QJ}. Then its time integral, 
\begin{eqnarray}
  P_{\omega_2} = \int\limits_0^{\infty}\text{d}t\, w_{1\omega_2}(t),
\end{eqnarray} 
is the probability for the next emitted photon to come from the
$|3\rangle - |2\rangle$ transition. Now, let the photon rate in a
light period be denoted by $I_L$. Then, after each photon of the light
period the system is reset to the ground state and thereafter, with
probability $P_{\omega_2}$, emits a photon from the $|3\rangle -
|2\rangle$ transition. Hence the transition rate from a light to a
dark period is 
\begin{equation}
p_{10} = I_L P_{\omega_2}.
\end{equation}
This can be carried over to two dipole interacting $D$ systems as
follows. We consider an emission trajectory and assume to be in a
particular intensity period, of unit intensity, say. In contrast to a
single $D$ system, the reset state after a photon emission in this
period is not always quite the same, but it is reasonable to start
from $\rho_{\text{ss},1}$ and to use 
\begin{eqnarray}
{\overline\rho}_1 & \equiv & \Big\{ (A_3 + \text{Re} C_3)R_+^{(3)}
    \rho_{\text{ss},1} R_+^{(3)^{\scriptstyle \dagger}} \nonumber \\
    \hspace{-1cm} && {}+
    (A_3-\text{Re} C_3) R_-^{(3)} \rho_{\text{ss},1}
    R_-^{(3)^{\scriptstyle \dagger}}\Big\}/{\rm Tr}(\cdot) 
\end{eqnarray}
as an average reset state. The transition to a double-intensity period
is marked by a photon from the $|2\rangle -|1\rangle$ transition, and
therefore the probability density for such a transition, starting
from the above reset state, is 
\begin{multline}
  w_{1\omega_1}(t) = \\ 
  \text{Tr}\Big\{(A_1+\text{Re}\,C_1)R_+^{(1)}\text{e}^{-i
    H_{\text{cond}}t/\hbar}{\overline\rho}_1 \text{e}^{i
    H^{\scriptstyle \dagger}_{\text{cond}}t/\hbar}R_+^{(1)^{\scriptstyle
    \dagger}} \\ {}+
(A_1-\text{Re}\,C_1)R_-^{(1)}\text{e}^{-i
  H_{\text{cond}}t/\hbar}{\overline\rho}_1 \text{e}^{i H^{\scriptstyle
    \dagger}_{\text{cond}}t/\hbar} R_-^{(1)^{\scriptstyle \dagger}}\Big\} 
\end{multline}
Integration over $t$ gives the total transition probability, denoted
by $P_{1\omega_1}$. The photon rate in a period of unit intensity is that
of two dipole interacting two level systems and is given by
\cite{BeHe:98}
\begin{widetext}
\begin{equation}
  I_{\text{ss}}^{(2)}= 2\frac{\Omega_3^2[A_3(A_3^2 + 2\Omega_3^2 +
    4\Delta_3^2) + \text{Re}\,C_3(A_3^2 + 4\Delta_3^2)]}{(A_3^2 + 2\Omega_3^2 +
    4\Delta_3^2)^2 + (A_3^2 + 4\Delta_3^2) (|C_3|^2 + 2A_3\text{Re}\,C_3 + 4\Delta_3\text{Im}\,C_3)}.
\end{equation}
\end{widetext}
Thus $p_{12}$ is given by
\begin{equation}
p_{12} = I^{(2)}_{ss} P_{1\omega_1}.
\end{equation}
In a similar way one obtains $p_{10}$ and $p_{21}$. The transition
rate $p_{01}$ can be directly read off from the no-photon probability
$e^{-2A_1t}$. One obtains the same results as in Section \ref{2Dpert} when one
expands in the small parameters.
In the case $\lambda_1,\lambda_3 \ll r \ll \lambda_2$ one can put
$C_1=C_3=0$ and one obtains for example 
\begin{equation}
  p_{12}=A_1\left(1+ \frac{\text{Im}\,C_2^2}{A_3^2 +2\Omega_3^2 + 4\Delta_3^2}\right).
\end{equation}


\begin{thebibliography}{27}
\bibitem[{rem()}]{remark}
\bibinfo{note}{In Ref. \cite{SaBlNeTo:86} it is stated ``We have also
  demonstrated, with the light of two or three ions in the trap observed, that
  simultaneous quantum jumps use to happen at a rate large exceeding random
  coincidence.''}.
\end{thebibliography}
\end{document}